\documentclass[aps,prl,twocolumn,superscriptaddress,showpacs]{revtex4}
\usepackage{graphicx}
\usepackage{epsfig}
\usepackage{color}
\usepackage[latin1]{inputenc}
\usepackage{amsmath}
\usepackage{amssymb}
\input{epsf}

\newcommand{\rem}[1]{}

\newcommand{\TT}{\mathcal{T}_3}

\newcommand{\virg}[1]{``{#1}''}

\newcommand{\vett}[1]{\mathbf{#1}}

\begin{document}

\title{$4e$-condensation in a fully frustrated Josephson junction diamond chain}

\author{Matteo Rizzi}
\affiliation{NEST CNR-INFM $\&$ Scuola Normale Superiore, Piazza
      dei Cavalieri 7, 56126 Pisa, Italy}

\author{Vittorio Cataudella}
\affiliation{COHERENTIA CNR-INFM $\&$ Dipartimento di Fisica,
      Università Federico II, 80126 Napoli, Italy}

\author{Rosario Fazio}
\affiliation{NEST CNR-INFM $\&$ Scuola Normale Superiore, Piazza
      dei Cavalieri 7, 56126 Pisa, Italy}
\affiliation{International School for Advanced Studies (SISSA)
        via	 Beirut 2-4,  I-34014, Trieste - Italy}
\date{\today}

\begin{abstract}
Fully frustrated one-dimensional diamond Josephson chains have been
shown [B. Dou\c{c}ot and J. Vidal, Phys. Rev. Lett. {\bf 88}, 227005
(2002)] to posses a remarkable property: The superfluid phase occurs
through the condensation of pairs of Cooper pairs. By means of Monte
Carlo simulations we analyze quantitatively the Insulator to
$4e$-Superfluid transition. We determine the location of the critical
point and discuss the behaviour of the phase-phase correlators. For 
comparison we also present the case of a diamond chain at zero and 
$1/3$ frustration where the standard $2e$-condensation is observed.
\end{abstract}

\maketitle

Josephson arrays in the quantum regime have been studied
extensively~\cite{fazio01}, both experimentally and theoretically, as
model systems where to investigate a variety of quantum phase
transitions. The application of a magnetic field creates frustration
and leads to a number of interesting physical
effects\cite{fazio01,classical}.

Very recently, renewed interest in frustrated Josephson networks has 
been stimulated by the work by Vidal {\em et al.}~\cite{vidal98} on 
the existence of localization in fully frustrated tight binding models 
with $\TT$ symmetry. Localization in this case is due the destructive 
interference for paths circumventing every plaquette. These clusters 
over which localization takes place were named {\em Aharonov-Bohm} 
(AB) {\em cages}. Experiments in superconducting networks have been 
performed and the existence of the AB cages has been confirmed through 
critical current measurements both in wire~\cite{abilio99} and 
junction~\cite{serret03,martinoli05} networks. Starting from the 
original paper by Vidal {\em et al.} several aspects of the AB cages 
both for 
classical~\cite{korshunov02,korshunov01,cataudella03,korshunov03} and 
quantum~\cite{rizzi05} superconducting networks have been highlighted.

\begin{figure}
      \centerline{\includegraphics[width=8cm]{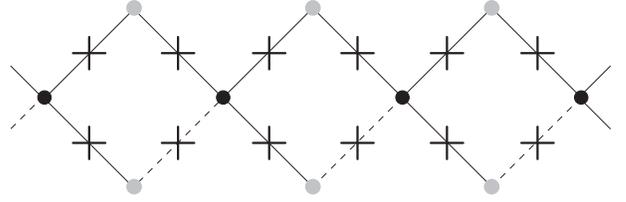}}
\caption{
   The diamond chain Josephson network analyzed in the present paper.
   The crosses represent the Josephson junctions connecting two
   neighboring superconducting islands. In the chain there are two types
   of inequivalent sites with two (grey) and four (black) neighbours. By an
   appropriate choice of the gauge, the magnetic phase factors $A_{i,j}$ can be
   chosen to be zero on the three links indicated by a continuous lines
   and $f$ in the fourth one indicated by a dotted line.}
\label{chain}
\end{figure}

The basic mechanism leading to the AB cages is also present
in the (simpler) quasi-one-dimensional lattice shown in
Fig.\ref{chain}. At fully frustration, it has been
shown~\cite{doucot02} that superconducting coherence is established
throughout the system by means of $4e$-condensation. The global
superconducting state is due to the condensation of {\em pairs} of
Cooper pairs. Predictions on the critical current of the diamond chain
of Fig.\ref{chain} amenable of experimental confirmation have been put
forward by Protopopov and Feigelman~\cite{protopopov04,protopopov05}. Unusual
transport properties of these systems have been also predicted in
semiconducting samples~\cite{bercioux04}.

In this work we present the results of our Monte Carlo simulations on
the Josephson junction network with the geometry depicted in
Fig.\ref{chain}. Our aim was to perform a detailed quantitative
analysis of the phase diagram predicted in Ref.\cite{doucot02}. In 
order to have a fairly complete description of the effect of 
frustration in this case, we considered the stiffness and phase 
correlators for three values of the frustration parameter; i.e. $f=0$, 
$f=1/3$ and $f=1/2$.

The Hamiltonian for a Josephson junction network is
\begin{equation}
        \mathcal{H} =
      E_0 \sum_{i} n_i^2 -E_J \sum_{\langle i,j \rangle}\cos (\varphi_i
      - \varphi_j - A_{i,j}) \;\;  .
\label{hamiltonian}
\end{equation}
The first term in the Hamiltonian is due to the charging energy. Here
for simplicity we consider the case in which the Coulomb interaction 
is on-site; see Ref.\cite{protopopov04} for the more realistic case of 
long range charging interaction. The second term is the Josephson 
contribution. The phase of the superconducting order parameter in the 
i-th island is denoted by $\varphi_i$, $E_0$ is the charging energy 
and $E_J$ is the Josephson coupling energy. The number $n_i$ and phase 
$\varphi_i$ operators are canonically conjugate on each site $
         \left[ n_{i}, e^{\imath
         \varphi_{j}} \right] = \delta_{ij} \,
         e^{ \imath \varphi_{i}}$.
The gauge-invariant definition of the phase in presence of an external 
vector potential $\vett{A}$ and flux-per-plaquette $\Phi$ ($\Phi_0 = h 
\, c / \, 2 \, e$ is the flux quantum) contains the term $ A_{i,j} = 
\frac{2 \pi}{\Phi_0} \int_{i}^{j} \vett{A} \cdot d\vett{l} \; . $ All 
the observables are function of the frustration parameter defined as
\begin{equation*}
         f = \frac{1}{\Phi_0}
         \int_{P} \vett{A} \cdot d\vett{l} \; = \frac{1}{2 \pi}
         \sum_{P} A_{i,j}
\end{equation*}
where the line integral is performed over the elementary plaquette.
Due to the periodicity of the model it is sufficient to consider values of
the frustration $0 \le f \le 1/2$.

The Monte Carlo simulations have been performed on an effective
classical $d+1$-dimensional XY-model~\cite{jacobs88,wallin94} (here
$d=1$) whose action is given by
\begin{eqnarray}
      \nonumber
      \mathcal{S} & = &
      - K
      \sum_{i, \langle k k' \rangle}
      \cos \left(\varphi_{i,k} -
      \varphi_{i,k'} \right) \\
      & &
      - K
      \sum_{\langle i j \rangle , k}
      \cos \left(\varphi_{i,k} -
      \varphi_{j,k} -
      A_{i,j}\right) \, .
\label{isotropic}
\end{eqnarray}
The effective dimensionless coupling is defined as $K=\sqrt{E_J/ E_0}$.
The index $k$ labels the extra (imaginary time) direction which takes
into account the quantum fluctuations (the vector potential does not 
depend on the imaginary time). The first term corresponds to charging 
while the second is due to the Josephson coupling. The simulations 
where performed on $L \times L $ lattice with periodic boundary 
conditions (the largest lattice was $72 \times 72$). In 
Eq.(\ref{isotropic}) the couplings along the time and space directions 
have been made equal by a proper choice of the Trotter time 
slice~\cite{jacobs88,wallin94}. This choice, with no consequences on 
the study of the zero temperature phase transition, makes the analysis 
of the Monte Carlo data considerably simpler. The expectation values 
of the different observables (stiffness and correlation functions) 
have been obtained averaging up to $10^{7}$ Monte Carlo configurations 
by using a standard Metropolis algorithm. Typically the first half of 
configurations, in each run, were used for thermalization. 

The stiffness, related to the critical current, is used to signal the
presence of the transition. It is defined through the increase of the
free energy ${\cal F}$ due to a phase twist $\delta $ imposed along
the space direction~\cite{ohta}:
\begin{equation*}
\Gamma =\frac{\partial ^{2}{\cal F}}{\partial \delta ^{2}}\;\;.
\end{equation*}
The critical point is expected to be of the Berezinskii-Kosterlitz-Thouless universality 
class~\cite{BKT,doucot02}. Its location can be determined using the following {\em ansatz} for
the size dependence of $\Gamma (K_c) $~\cite{weber}
\begin{equation}
\frac{\pi K_{c}}{2} \; \Gamma_L (K_c) = 1 + \frac{1}{2\ln (L/l_{0})}
\label{stiffansatz}
\end{equation}
where $l_{0}$ is the only fit parameter. In the presence of
frustration, the universality class of the transition may be different
from that of the unfrustrated case. In the case of the two-dimensional
fully frustrated XY-model this issue has been investigated in great
detail (see Ref. \onlinecite{olsson} and refs. therein). Up to
date, there is no unanimous consensus on the nature of the transition.
However, in this work we suppose that the transition belongs to the BKT
universality class, as suggested by Ref.\cite{doucot02}, and determine
the critical value by means of Eq.(\ref{stiffansatz}). 

\begin{figure}
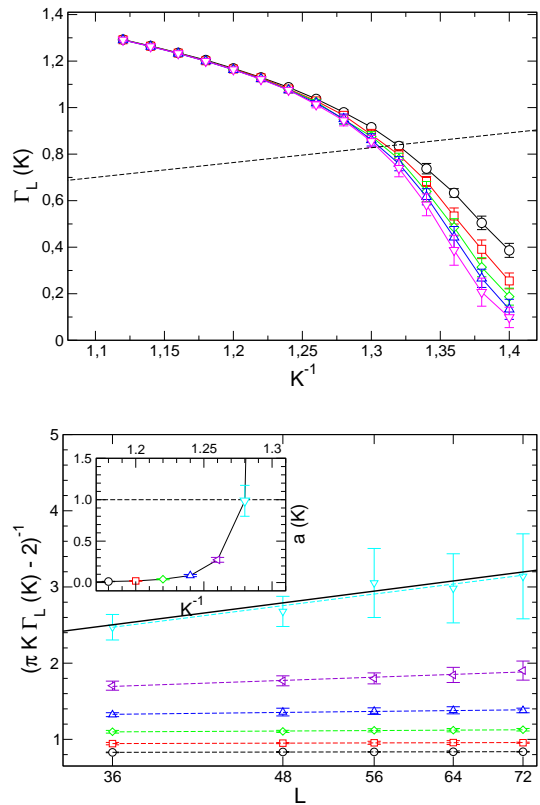

     \centerline{\includegraphics[width=7cm]{stiff_00}}
     \vspace{5mm}
     \centerline{\includegraphics[width=7cm]{trans_00_bis}}
\caption{
   In the upper panel, the stiffness for the case of $f=0$ frustration
   is plotted against the coupling. Different symbols correspond to
   different sizes of the chain: circles to $L=36$, squares to $L=48$,
   diamonds to $L=56$, triangles up to $L=64$, and triangles down to 
   $L=72$. The dashed line with slope $2/\pi$ gives a rough estimate of	 
   the transition point. A better estimate is obtained by means of the 
   finite size scaling shown in the lower panel and explained in the 
   text.	 Thick black line has exactly slope 1 and is plotted as a 
   reference guide. The value of $l_0$ at the transition is $2.9$}
\label{stiff00}
\end{figure}

We first analyze the $f=0$ case and extract the value of the critical 
coupling from the stiffness. This extrapolation has been done by 
performing a linear fit in logarithmic scale $\left[ \pi \, K \, 
\Gamma_L (K) - 2 \right]^{-1} = a(K) \ln L - \ln l_0$ and searching 
for the coupling value such that $a(K) = 1$. This coupling value is 
then identified with the critical point $K_c$. The proposed ansatz 
fits very well the data and the estimated value of the critical 
coupling is $K_c^{-1} = 1.28$ which corresponds to $(E_J/E_0)_c \sim 
0.61$. Data are reported in Fig.\ref{stiff00}.

\begin{figure}
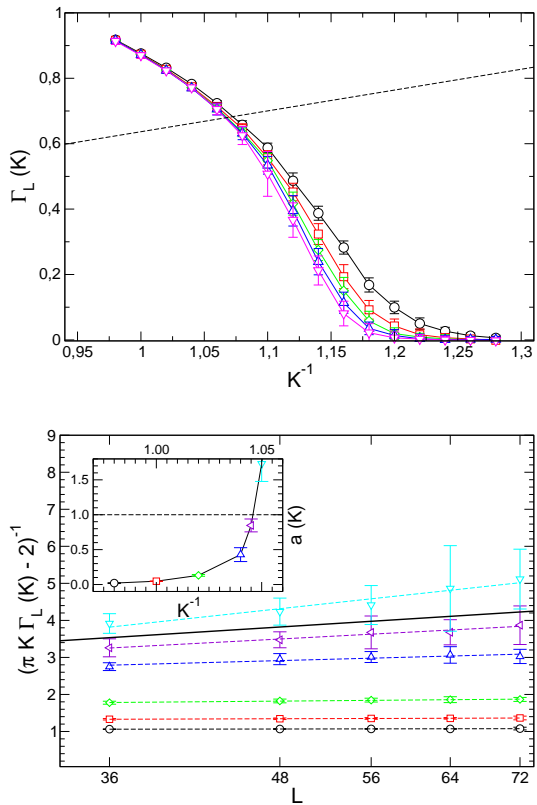

     \centerline{\includegraphics[width=7cm]{stiff_03}}
     \vspace{5mm}
     \centerline{\includegraphics[width=7cm]{trans_03_bis}}
\caption{
   The same plots of Fig.\ref{stiff00} for the case of $f=1/3$.
   The critical point is $K_c^{-1} = 1.045$ (with $l_0 \sim 0.6$).}
\label{stiff13}
\end{figure}

\begin{figure}
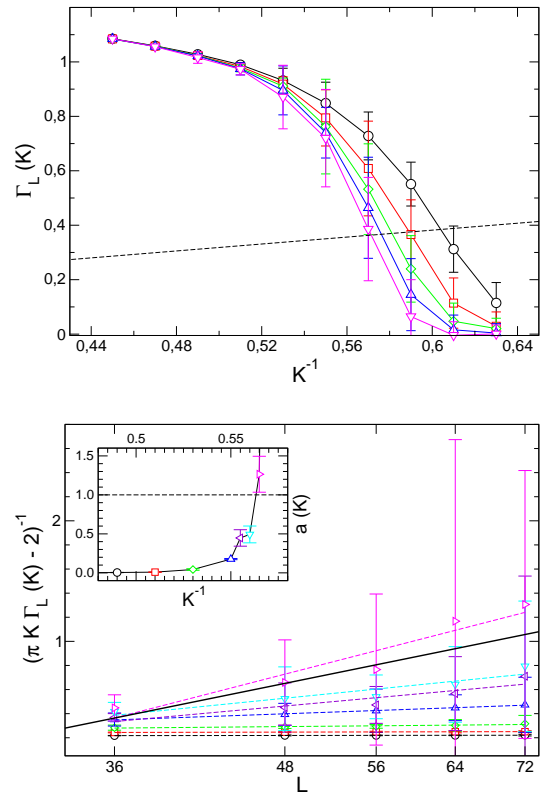

     \centerline{\includegraphics[width=7cm]{stiff_05}}
     \vspace{5mm}
     \centerline{\includegraphics[width=7cm]{trans_05_bis}}
\caption{
   The same plots of Figs.\ref{stiff00}-\ref{stiff13} for $f=1/2$.
   Compared to the cases of $f=0$ and $f=1/3$, the superfluid region is
   considerably shrunk. The value of $l_0$ at the transition is $\sim 25$.}
   \label{stiff12}
\end{figure}

\begin{figure}[hbtp]
     \centerline{\includegraphics[width=7cm]{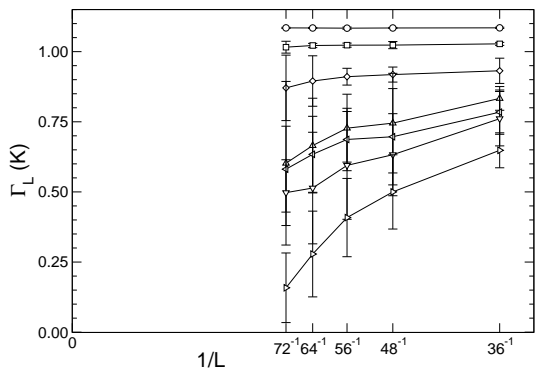}}
\caption{The stiffness is plotted as a function of the system size for 
   different values of $K$ in the critical region. This plot 
   highlights the existence of a transition, though do not allow to 
   extract the transition point. For $K^{-1} < 0.55$ data seem to scale 
   to a finite value in the thermodynamic limit, whereas over $0.57$ it 
   seems clear that they go to zero. Different symbols correspond to 
   values of $K^{-1}$: circles ($0.45$), squares ($0.49$), diamond 
   ($0.53$), triangles up ($0.555$), triangles left ($0.56$), triangles 
   down ($0.565$), triangles right ($0.575$).}
\label{fruscal}
\end{figure}

The results of the stiffness at $f=1/3$ and $1/2$ are shown in
Fig.\ref{stiff13}
and Fig.\ref{stiff12} respectively. As compared to the unfrustrated 
case, the critical value of the Josephson coupling required to 
establish superfluid coherence is slightly larger for $f=1/3$ and 
further increases for the fully frustrated case $f=1/2$. The ansatz of 
Eq.~(\ref{stiffansatz}) seems to provide an accurate estimate of the 
transition point for $f=0$ and $f=1/3$. In the fully frustrated case, 
however, the value of $l_0=25$ indicates that we probably need larger 
chains in order to really enter the critical region. Another 
indication of this fact emerges in the upper panel of 
Fig.\ref{stiff12} where the line of slope $2/\pi$ crosses the data 
already when the stiffness is decreasing to zero. In order to put 
bounds to the critical point in the fully frustrated case we plot in 
Fig.\ref{fruscal} the stiffness as a function of the system size.
From the raw data it is possible to bound the transition point in the 
range $0.55 \le K_c^{-1} \le 0.57$. 

All these results are summarized in table below:
\begin{center}
\begin{tabular}{|c|c|c|c|}
   \hline
   $f$ & $0 $ & $1/3$ & $1/2$ \\
   \hline
   $K_c^{-1}$ & $1.28 \pm 0.01$ & $1.045 \pm 0.005$ & $0.56 \pm 0.01$ \\
   $(E_J/E_0)_c$ & $0.61 \pm 0.01$ & $0.91 \pm 0.01$ & $3.2 \pm 0.1$ \\
   \hline
\end{tabular}
\end{center}
The ratio of the obtained critical couplings for unfrustrated and 
fully frustrated system is $K_{c,1/2}/K_{c,0}=2.28 \pm 0.06$ and not 
$4$ as expected from the reduction by a factor $1/2$ of the effective 
charge of the topological excitation that unbind at the critical 
point. This may be due to the fact that the screening of the vortices 
is different in the unfrustrated and fully frustrated case therefore 
leading to a further correction in the ratio between the two critical 
points.

\begin{figure}
     \centerline{\includegraphics[width=9cm]{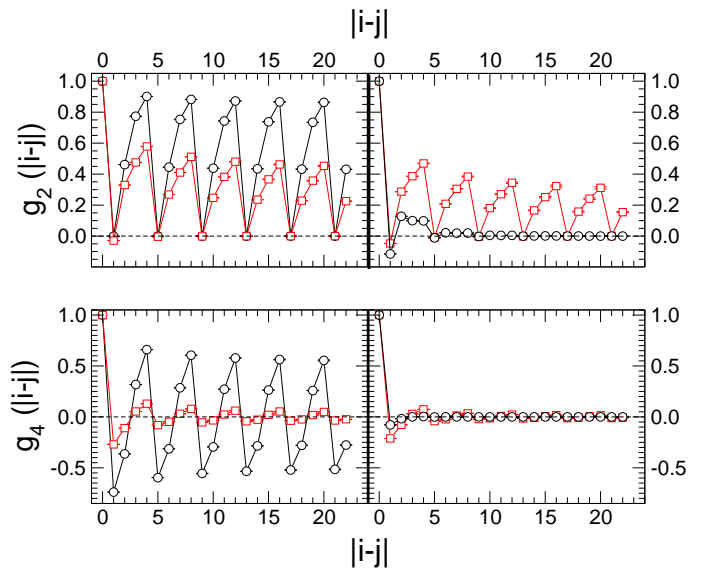}}
\caption{
   Phase correlators at frustration $f=1/3$.
   Up: the phase correlator $g_2(|i-j|)$ is shown as a function of the
   distance between the sites both in the ordered phase (left panel),
   $K^{-1}= 0.3$ (circles) and $1.0$ (squares) and in the Mott insulator
   phase (on the right) $K^{-1}= 1.1$ (squares) and $1.4$ (circles).
   Differently from the fully frustrated case, here the phase correlator
   of Cooper pairs changes its behaviour at the critical point.
   Down: the phase correlator $g_4(|i-j|)$ is shown for the same
   coupling values as upwards.}
\label{corr2f13}
\end{figure}

The differences in the fully frustrated case manifest dramatically in the
way condensation is achieved. As predicted by Dou\c{c}ot and
Vidal~\cite{doucot02}, the destructive interference built in the
diamond structure prevents Cooper pair to have (quasi-)long range
order. The superfluid phase is then established via the delocalization
of pairs of Cooper pairs. This is at the origin of the
$4e-$condensation. In order to check this point, the knowledge of the
phase-phase correlators is required. Quasi-long range behaviour in a
two-point correlation function of the type
\begin{equation}
g_{2n}(|i-j|) = \langle \cos n(\varphi_i - \varphi_j)\rangle
\end{equation}
signals the existence of condensation of $2n$ charged objects.
\begin{figure}
     \centerline{\includegraphics[width=9cm]{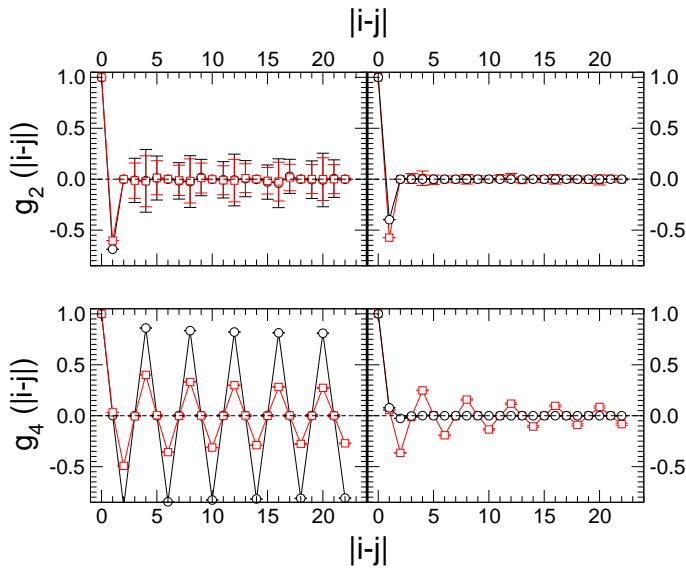}}
\caption{
   The phase correlators $g_2$ and $g_4$ are shown as a function of the
   distance for the \emph{fully frustrated} case $f=1/2$. Data are
   plotted for a chain with $L=48$.
   On the left side circles correspond to $K^{-1}=0.1$ deep in the
   ordered phase and squares to $K^{-1}=0.5$ on the border of it. On the
   right side, squares are $K^{-1}=0.6$ and circles $K^{-1}=1.2$, deep
   in the Mott insulator phase.
   Differently from $g_2$, the correlator $g_4$ shows quasi-long range
   order.} \label{corr2}
\end{figure}
In Fig.\ref{corr2} we discuss their properties. In the upper panels,
we consider the phase-phase correlator $g_2$ for two different
couplings deep in the superfluid and Mott insulating phases
respectively. What is evident from the figure is that, despite the
fact that the system is phase coherent, phase correlations decay very
fast almost independently from the value of $K$. As explained
in~\cite{doucot02}, this behaviour should be ascribed to the existence
of the Aharonov-Bohm cages. Even if hopping of single Cooper pairs is
forbidden because of quantum interference, correlated hopping of two
pairs does not suffer the same destructive interference. In the lower
panels of the same figure, the space dependence of the correlator
$g_4$ is plotted for the same coupling as upwards. The different
behaviour between the Mott and the superfluid phase is now evident.
The correlator decays exponentially only for $K^{-1} = 1.2 > K_c$
(right side): in the other panel, differently from $g_2$, the decay is
power-law like.
For comparison we report also simulations of the phase correlators for
the case $f=1/3$. In this case the \virg{standard} condensation of
Cooper pair is observed as witnessed by the behaviour of $g_2$ shown in
Fig.\ref{corr2f13}.

We acknowledge useful discussions with Michele Governale and Julien
Vidal. This work has been supported by MIUR-FIRB and by EC Community
through contracts RTNNANO and SQUBIT2.

\bibliographystyle{prsty}

\begin{thebibliography}{99}
\bibitem{fazio01}
      R. Fazio and H.S.J. van der Zant,
      Phys. Rep. {\bf 355}, 235 (2001).
\bibitem{classical}
         R.S. Newrock, C.J. Lobb, U. Geigenmller,
         and M. Octavio, Solid State Phys. {\bf 54}, 263 (2000).
\bibitem{vidal98}
        J. Vidal, R. Mosseri, and B. Dou\c{c}ot,
        Phys. Rev. Lett. {\bf 81}, 5888 (1998).
\bibitem{abilio99}
        C. Abilio, P. Butaud, Th. Fournier, B. Pannetier,
        J. Vidal, S. Tedesco, and B. Dalzotto,
      Phys. Rev. Lett. {\bf 83}, 5102 (1999).
\bibitem{serret03}
        E. Serret,  P. Butaud, and B. Pannetier,
        Europhys. Lett. {\bf 59}, 225 (2003).
\bibitem{martinoli05}
        M. Tesei, R. Th\'{e}ron, and P.Martinoli, cond-mat/0510033
        (2005).
\bibitem{korshunov02}
        S.E. Korshunov, Phys. Rev. B {\bf 65}, 054416 (2002).
\bibitem{korshunov01}
        S.E. Korshunov, Phys. Rev. B {\bf 63}, 134503 (2001)
\bibitem{cataudella03}
        V. Cataudella, and R. Fazio, Europhys. Lett. {\bf 61}, 341 (2003).
\bibitem{korshunov03}
        S.E. Korshunov and B. Dou\c{c}ot, Phys. Rev. Lett. {\bf 93}, 097003
        (2004).
\bibitem{rizzi05}
        M. Rizzi, V. Cataudella, and R. Fazio, cond-mat/0510341
\bibitem{doucot02}
        B. Dou\c{c}ot, and J. Vidal,
        Phys. Rev. Lett. {\bf 88}, 227005 (2002).
\bibitem{protopopov04}
        I.V. Protopopov, and M.V. Feigel'man, Phys. Rev. B \textbf{70},
      184519 (2004).
\bibitem{protopopov05}
        I.V. Protopopov, and M.V. Feigel'man,
   cond-mat/0510766
\bibitem{bercioux04}
         D. Bercioux, M. Governale, V. Cataudella, and V.
         M. Ramaglia, Phys. Rev. Lett. {\bf 93}, 56802 (2004).
\bibitem{jacobs88}
         L. Jacobs, J. V. José, M. A. Novotny, and A. M. Goldman
         Phys. Rev. B {\bf 38}, 4562 (1988).
\bibitem{wallin94}
         M. Wallin, E.S. S\o rensen, S. M. Girvin, and A. P. Young,
         Phys. Rev. B {\bf 49}, 12115 (1994).
\bibitem{ohta}
      T.~Ohta, and D.~Jasnow, Phys. Rev. B {\bf 20}, 139 (1979).
\bibitem{weber}
      H.~Weber, and P.~Minnhagen, Phys. Rev. B {\bf 37}, 5986
        (1988).
\bibitem{BKT}
      V.L.~Berezinskii, Zh. Eksp. Teor. Fiz. {\bf 59}, 207 (1970)
        [Sov. Phys. JETP {\bf 32}, 493 (1971)]; J.M.~Kosterlitz and
        D.J.~Thouless, J. Phys. C {\bf 6}, 1181 (1973).
\bibitem{olsson}
        P.~Olsson, Phys. Rev. B {\bf 55}, 3585 (1997).



\bibitem{fubini}
   K.~Harada, and N.~Kawashima, J. Phys. Soc. Jpn. {\bf 67}, 2768
   (1998); A.~Cuccoli, T.~Roscilde, V.~Tognetti, R.~Vaia, P.~Verrucchi,
   Phys. Rev. B, {\bf 67}, 104414 (2003); L.~Capriotti, A.~Cuccoli,
   A.~Fubini, V.~Tognetti, R.~Vaia, Phys.Rev.Lett. 94, 157001 (2005)
\end{thebibliography}

\end{document}